\documentclass[10pt, aip,reprint]{revtex4-1}

\draft 
\usepackage{hyperref}
\usepackage{amsmath, amssymb, graphicx, xcolor}
\usepackage{dcolumn}
\usepackage{bm}

\usepackage[utf8]{inputenc}
\usepackage[T1]{fontenc}
\usepackage{mathptmx}
\usepackage{etoolbox}
\graphicspath{{~/Nextcloud/Projects/Alladi_Erdos/Figures/}}

\begin{document}


\title{Attractors and basins generated by repeated sums of prime factors of natural numbers}



\author{Snehal Shekatkar}
\homepage[]{https://inferred.in}
\email[]{snehal@inferred.in}
\affiliation{School of Computing and Data Sciences, FLAME University, Pune 412115, India}

\author{Ivano Lodato}
\email[]{ivano.lodato@allos.ai}
\affiliation{Allos AI Limited, Hatton Garden, London, EC1 N8LE, United Kingdom}

\begin{abstract}
Integer maps are discrete dynamical systems defined on the natural numbers. In this paper, we investigate the dynamics of the shifted Alladi-Erd{\H o}s map, a one-parameter family of integer maps in which each composite number is mapped to the sum of its prime factors, while each prime is mapped to $n+A$, where $A \in \mathbb{N}$ is a fixed shift parameter. By systematically exploring the parameter space for $2 \leq A \leq 10^5$, we uncover a rich bifurcation structure characterized by the emergence, disappearance, and reorganization of attractor cycles as the shift parameter varies. We find that although several attractors may coexist for a given value of $A$, for most values of $A$, almost all natural numbers belong to the basins of only two dominant attractors. Using elementary number theoretic arguments, we explain the observed bifurcation diagram. We further characterize the attractor cycles and quantify the distribution of their basin sizes across the parameter space. Our results reveal an unexpectedly rich landscape of arithmetic dynamics arising from a remarkably simple integer map. 
\end{abstract}


\maketitle 

\begin{quotation}
The \emph{shifted} Alladi-Erd{\H o}s map is a discrete-time dynamical system defined on natural numbers greater than $1$. For a given natural number $n$, if it is composite, the map sends it to the sum of its prime factors counted with multiplicity, whereas if it is prime, it is sent to $n+A$ where $A$ is called ``shift''. For different values of $A$, map generates a different set of attractors, which are sets of numbers on which all initial conditions ultimately settle. All the attractors of the map (except number $4$ which is the only fixed point) are cycles whose number of sizes depend on the value of the shift parameter. Here we present results of numerical computation with shift parameter varied up to $10^5$ which reveals unseen properties of the map in terms of its attractor structure and their basins. In particular, we find that for most values of $A$, just two cycles attract most of the initial conditions to them. The map illuminates the relationship between numbers and their prime factorization. 
\end{quotation}
\section{Introduction}
\label{sec:introduction}
Iterated maps on the integers are among the simplest systems capable of producing complicated dynamics. Well-known examples include the Collatz map \cite{lagarias1985collatz}, aliquot sequences \cite{guy1975aliquot}, and maps constructed from digits \cite{elsedy2000happy} or divisor functions \cite{cohen1996divisors}. Unfortunately, many geometric tools used for smooth dynamical systems are unusable for these systems as their phase spaces are discrete and arithmetically structured. Nevertheless, the usual dynamical questions like which attractors exist, their coexistence and basins, and their parameter dependence, remain meaningful. 

In this paper, we study the shifted Alladi-Erd{\H o}s map introduced in Ref.~\cite{shekatkar2017shifts} whose phase space consists of all natural numbers $n>1$. The map sends a natural number $n$ to the sum of its prime factors, counted with multiplicity, when $n$ is composite, and to $n+A$ when $n$ is prime. Here the natural number $A$ is called the shift parameter. Previous work proved that every trajectory of the map eventually becomes periodic, and that there are no unbounded trajectories~\cite{shekatkar2017shifts}.

However, the fact that there are no unbounded trajectories does not imply that the map's attractors and basins have a simple structure. The numerical exploration in \cite{shekatkar2017shifts}, although restricted to a small range  $1 \leq A\leq 200$, already showed that the number of attractors seems to vary quite irregularly with $A$, which raises an immediate question about the behavior of the map over a larger range of shifts. Moreover, the work did not examine the basin structure of the map at all. In this work, we try to fill this gap by performing a substantially larger computational study, reaching shifts up to $A=10^5$ and initial conditions up to $N=10^6$. We treat $A$ as a discrete control parameter and ask how the global dynamics changes as it is varied. Three observations organize the paper. First, apart from the universal fixed point $4$, the sets of nontrivial attractors associated with distinct values of $A$ are disjoint. Thus changing the shift replaces every nontrivial attractor. Second, this complete turnover does not produce a featureless collection: the individual elements of attractors form clear line-like families as $A$ varies, and we analytically prove their existence using elementary number theory. Third, although several attractors may coexist, their basin sizes are strongly unequal, and for most values of $A$ the two largest basins account for almost all the initial conditions.

The rest of the paper is organized as follows. In Section~\ref{sec:map} we define the map and describe some of its simple properties including the existence of prime trees which are structures in the phase space of the map that are independent of the parameter $A$. In Section~\ref{sec:compressed} we study the attractor landscape across the shift parameter and  examine the number and lengths of coexisting attractors. In Section~\ref{sec:basins} we study the basin sizes of attractors and quantify them. Finally, Section~\ref{sec:discussion} connects the change in dynamics as $A$ is varied to bifurcations in dynamical systems, and summarizes the main conclusions. Computational details are described in Appendix~\ref{app:numerics}.

\section{The shifted map and its reduction to prime dynamics}
\label{sec:map}

Let
\begin{equation}
 n=\prod_{i=1}^{k}p_i^{r_i}
\end{equation}
be the prime factorization of $n$. The Alladi-Erd{\H o}s function is then defined as \cite{alladi1977additive}:
\begin{equation}
 B(n)=\sum_{i=1}^{k}r_i p_i,
 \label{eq:B}
\end{equation}
where prime factors are counted with multiplicity. The shifted map is defined for $n>1$ by
\begin{equation}
 B_A(n)=
 \begin{cases}
 n+A, & n \text{ prime},\\
 B(n), & n \text{ composite}.
 \end{cases}
 \label{eq:BA}
\end{equation}
Starting from an initial value $n$, repeated application of $B_A$ generates its forward trajectory. Previous work established that every such trajectory eventually enters a periodic cycle~\cite{shekatkar2017shifts} forming an attractor. The basin of an attractor consists of all initial values whose trajectories eventually enter that cycle.

We also note that the number $4$ is the only fixed point of the map independent of the value of $A$ as for every shift $B_A(4)=2+2=4$. It will thus be called the universal fixed point. Unless stated otherwise, counts of attractors below refer to nontrivial attractors, with $(4)$ excluded.

\begin{figure*}[t]
    \centering

    \begin{minipage}[t]{0.48\textwidth}
        \centering
        \includegraphics[width=\linewidth]{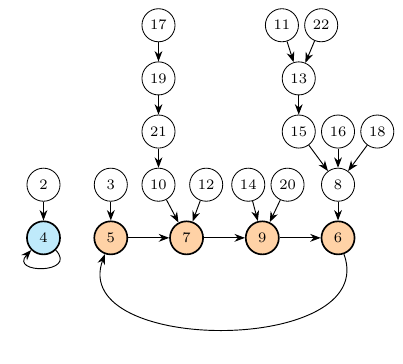}

        \smallskip
        \textbf{(a)} \(A=2\)
    \end{minipage}
    \hfill
    \begin{minipage}[t]{0.48\textwidth}
        \centering
        \includegraphics[width=\linewidth]{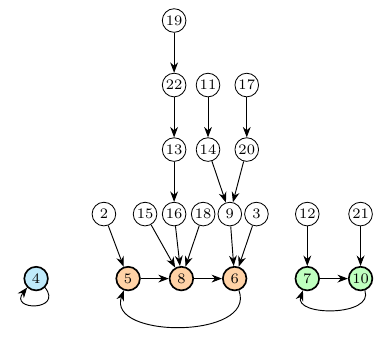}

        \smallskip
        \textbf{(b)} \(A=3\)
    \end{minipage}

    \caption{Trajectories of the shifted Alladi-Erd{\H o}s map for initial values \(2,\ldots,22\) for
    \(A=2\) and \(A=3\). Colored nodes indicate attractors. }
    \label{fig:trajectories}
\end{figure*}

Furthemore, for every composite $n>4$, one has $B(n)<n$. This simple inequality has two useful consequences. First, it implies that the smallest element of any nontrivial cycle must be prime because if the smallest element were composite and larger than $4$, its next iterate would be smaller, contradicting minimality. 
A consequence of this is that the numbers $2$ ad $3$ will never appear in any cycle: if $2$ could appear it would be the smallest prime of the cycle by definition, but there exists no composite whose prime factor decomposition gives $2$. The same argument applies to $3$. Since $n=4$ is a fixed point of the map, all non-trivial attractors consist of numbers greater than $4$.  
Second, the same nontrivial directed cycle cannot occur for two different shifts: if $q$ is the smallest element of such a cycle, then $q$ is prime and its successor is $q+A$. Fig.~\ref{fig:trajectories} shows trajectories of the map for initial conditions $2 \leq n \leq 22$ for $A=2$ and $A=3$.

\begin{figure*}[t]
    \centering
    \includegraphics[width=0.95\columnwidth]{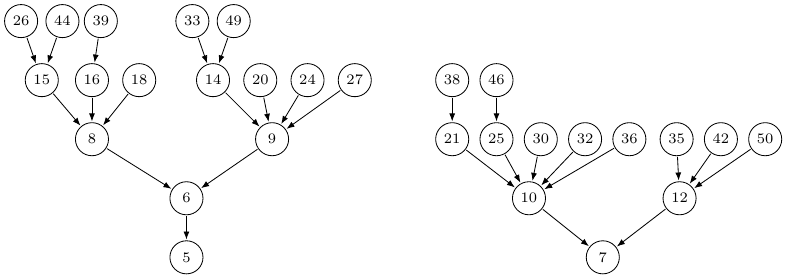}
    \caption{\label{fig:prime_trees}
    Schematics of prime trees of $5$ and $7$ containing all numbers $5 \leq n \leq 50$ that belong to either of the trees.} 
\end{figure*}

Interestingly, the inequality $B(n)<n$ for composite $n>4$ also reveals a useful decomposition of the phase space. If the unshifted function $B$ is applied repeatedly to a composite integer greater than $4$, the resulting decreasing trajectory eventually reaches either a prime. All integers that terminate at the same prime form a rooted directed tree, with edges pointing toward the root. We refer to these objects as \emph{prime trees}. The trees are determined entirely by $B$, and are therefore the same for every value of the shift parameter. Fig.~\ref{fig:prime_trees} shows the prime trees rooted at $5$ and $7$ with numbers less than $50$ included.

For each shift $A$, let $K_A$ be the number of nontrivial attractors. For an attractor $C$, we denote its full cycle length by $\ell(C)$ and its smallest element by $q(C)$ which, by the argument above, is a prime greater than or equal to $5$ for all $A$.

\section{Attractor landscape across the shift parameter}
\label{sec:compressed}

\begin{figure}[t]
    \includegraphics[width=\linewidth]{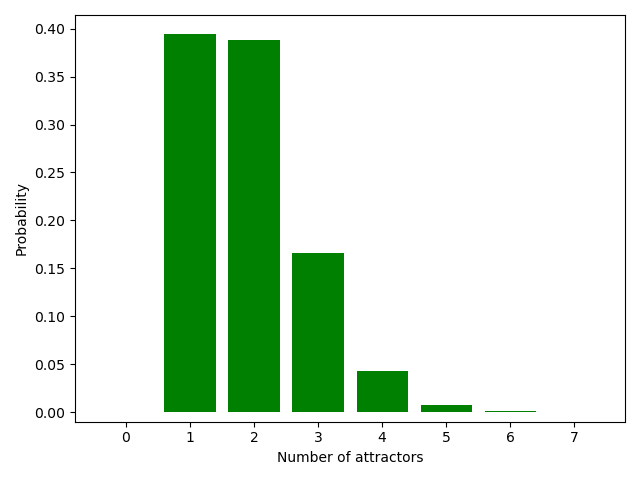}
    \caption{Histogram of the number of nontrivial attractors obtained by iterating the map up to $A=10^5$ with $N=10^6$, showing the probability to have a defined number of attractors.}
    \label{fig:num_cycles}
\end{figure}

We now move on to the question of quantifying the attractors as $A$ is varied. First let us look at the number of attractors for each value of $A$. We find that the number of nontrivial attractors changes extremely slowly: for $A$ up to $10^5$, the maximum number of nontrivial attractors is just $7$.
Fig.~\ref{fig:num_cycles} shows the histogram of the number of cycles for the range $A \in [1, 10^5]$.

\begin{figure}[t]
    \centering
    \includegraphics[width=0.8\columnwidth]{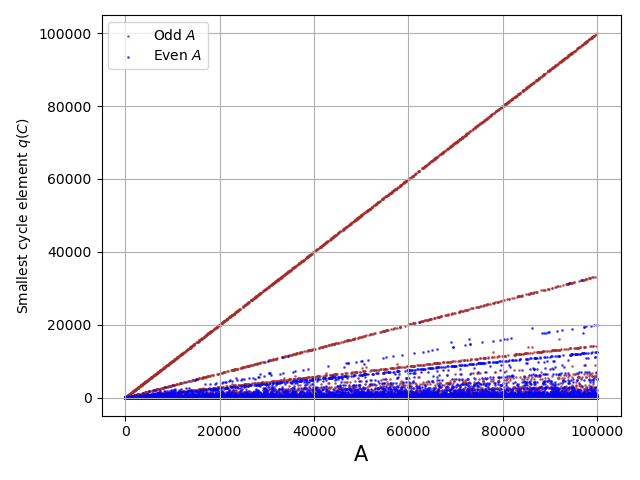}
    \caption{\label{fig:compressed}
    Compressed attractor landscape obtained by varying $A$ up to $10^5$ with $N=10^6$. For every nontrivial attractor at shift $A$, only its smallest element $q(C)$, which is necessarily a prime, is plotted. A prominent near-diagonal branch and several lower-slope branches are visible above a dense set of smaller minima.}
\end{figure}

Conventionally, the changes to the set of attractors in a dynamical system are visualized by plotting its bifurcation diagram as a parameter is varied. However, for the shifted Alladi-Erd{\H o}s map each value of the shift parameter $A$ has its own set of nontrivial attractors. Because of this a conventional bifurcation diagram quickly becomes crowded. 
Thus, although the full attractor diagram might useful for particular purposes, in this case it is not the clearest way to show the parameter dependence. To get over this problem, we use a ``compressed'' version of each attractors by using a single representative member of each attractor, and the smallest prime $q(C)$ is the most natural choice as the representative. Figure~\ref{fig:compressed} plots one point $q(C)$ for every nontrivial attractor while $A$ is the independent parameter on the horizontal axis. The most conspicuous feature of this plot is the branch close to the diagonal, accompanied by several other line-like structures. This shows that complete turnover of the individual cycles does not mean that maps structure changes completely irregularly as $A$ is varied. We explain this hidden order in the way attractors disappear and new ones appear later in this section using number theoretic arguments.

\begin{figure*}[t]
    \centering

    \begin{minipage}[t]{0.48\textwidth}
        \centering
        \includegraphics[width=\linewidth]{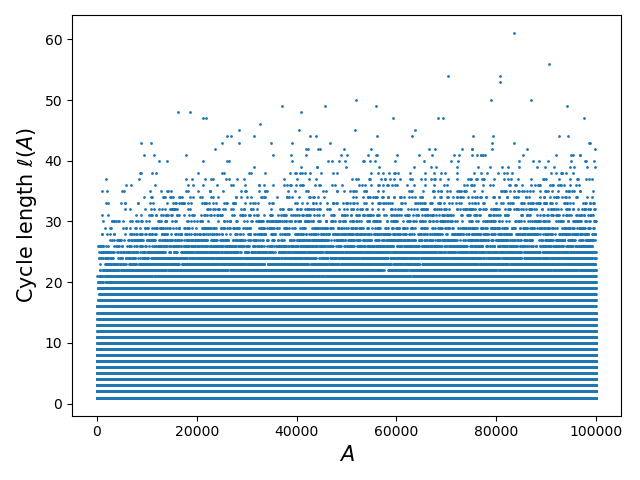}

        \smallskip
        \textbf{(a)} 
    \end{minipage}
    \hfill
    \begin{minipage}[t]{0.48\textwidth}
        \centering
        \includegraphics[width=\linewidth]{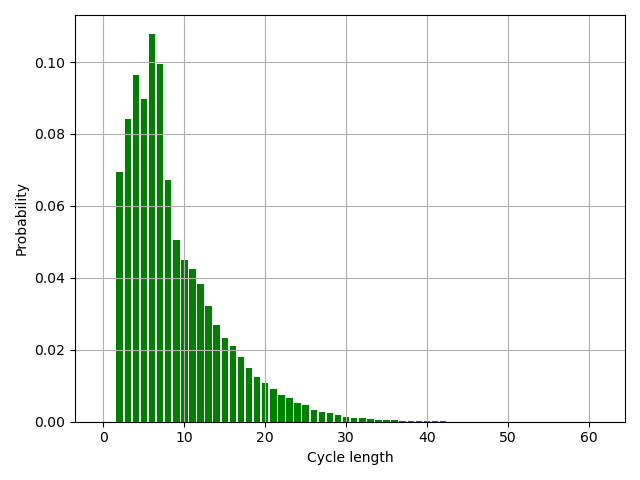}

        \smallskip
        \textbf{(b)} 
    \end{minipage}

    \caption{Properties of the attractor-cycle lengths for $1 \leq A \leq 10^5$. (a) Lengths of all attractor cycles found for each value of $A$. Multiple points at the same value of $A$ correspond to coexisting attractors. (b) Empirical distribution of the cycle lengths over the same range of $A$.}
    \label{fig:cycle_lengths}
\end{figure*}

Also, for completeness, the conventional diagram containing all attractor elements is shown in Appendix~\ref{app:full_diagram}. Its comparison with Fig.~\ref{fig:compressed} demonstrates why the minimum-prime representation is more effective for displaying structure across the parameter family while retaining the same line-like organization for the cycle elements vs $A$.
Fig.~\ref{fig:cycle_lengths}(a) shows the lengths $\ell(C)$ of attractors as a function of $A$, and part ($b$) shows the histogram of cycle lengths. 

We now prove the existence of the straight lines seen in Fig.~\ref{fig:compressed}. 
Let us consider a cycle of arbitrary length, with ordered elements $(n_1,n_2,\dots n_L)$ and $n_{L+1}=n_1$ where $n_1=p_1$ is the smallest element of the cycle, necessarily a prime. \\
The next step produces $n_2$ via the shift:
\begin{equation}
    n_1=p_1\rightarrow n_2=p_1+A
\end{equation}
Now there are two options: if $n_2=p_1+A=p_2$ prime, then the third element of the cycle will be $n_3=p_2+A=p_1+2A$.\\
If $n_2=C_1$ is the first composite number of the cycle starting from $p_1$, we can write it as the product of a prime/composite number $c_1$ and a prime number $q_1$:
\begin{equation}
    n_2=C_1=p_1+A=c_1 \cdot q_1 \;\; \Rightarrow q_1=\frac{n_2}{c_1}=\frac{p_1+A}{c_1}
\end{equation}
Now the next element $n_3$ reads:
\begin{equation}
 n_3=B(c_1)+q_1 = B(c_1)+ \frac{n_2}{c_1}= B(c_1)+\frac{p_1+A}{c_1}
\end{equation}
In general any step from any element of the cycle $n_i$ to the next will read:
\begin{equation}
n_{i+1}=\alpha_i n_i +\beta_i A + \gamma_i
\end{equation}
with 
\begin{align}
    \alpha_i&= \frac{1}{c_i} \;\;,\;\; \beta_i=0\;\;,\;\; \gamma_i=B(c_i) \qquad \qquad ({\rm for\;} n_i {\rm \;composite} )\nonumber\\
     \alpha_i&= 1 \;\;\;,\;\;\;\, \beta_i=1\;\;\;,\;\, \gamma_i=0 \qquad \qquad\qquad ({\rm for\;} n_i {\rm \; prime} )\nonumber 
\end{align}
So we have:
\begin{align}
 n_2 &= \alpha_1 n_1 + \beta_1 A + \gamma_1 \\
 n_3&=\alpha_2 n_2 + \beta_2 A + \gamma_2 = \alpha_2\alpha_1 n_1 + (\alpha_2 \beta_1 +\beta_2) A + (\alpha_2 \gamma_1 +\gamma_2)\\
 n_4&= \alpha_3\alpha_2\alpha_1 n_1 + (\alpha_3 \alpha_2\beta_1 +\alpha_3\beta_2 +\beta_3) A + (\alpha_3 \alpha_2\gamma_1 +\alpha_3\gamma_2 +\gamma_3)
\end{align}
which leads to the expression for the $L$-th element of the cycle in terms of the first $n_1=p$: 
\begin{equation}
    n_{L+1}= \left[\prod_{i=1}^L \alpha_i \right] n_1 + \left[\sum_{i=1}^L \beta_i \cdot \left(\prod_{j=i+1}^L \alpha_j\right) \right] A +  \left[\sum_{i=1}^L \gamma_i \cdot\prod_{j=i+1}^L\alpha_j\right]
    \end{equation}
which is a linear equation in $A$.\\
Now, because we are closing the cycle $n_{L+1}=n_1=p$, we can obtain an expression for $n_1=p_1$ in terms of $A$, i.e.:
\begin{equation}
    p_1=\left[\frac{\sum_{i=1}^L \beta_i \left(\prod_{j=i+1}^L \alpha_j\right)}{1-\prod_{i=1}^L \alpha_i}\right]A +\left[\frac{\sum_{i=1}^L \gamma_i \left(\prod_{j=i+1}^L\alpha_j\right)}{1-\prod_{i=1}^L \alpha_i}\right]
\end{equation}
This is an equation of a straight line which explains the straight lines in Fig.~\ref{fig:compressed}. Using a similar argument, it can be proved that any element of the cycle follows its own straight line which explains why the plot in Fig.~\ref{fig:full_bifurcation} contains so many straight lines. Properties of the straight lines and cycles will be analyzed in-depth in a follow-up paper. 

\section{Hierarchy and diversity of basin sizes}
\label{sec:basins}

In this section we discuss the basin structure of the cycles.
Basin sizes are measured within the finite window $2\leq n\leq N$. If $b_j(A;N)$ is the number of initial conditions that eventually enter attractor $j$, its basin fraction is
\begin{equation}
 p_j(A;N)=\frac{b_j(A;N)}{N-1}.
 \label{eq:basin_fraction}
\end{equation}
Notice that the denominator is $N-1$ instead of $N$ because the map is not defined for $n=1$, and hence the total number of initial conditions up to $N$ is reduced by $1$.
For each $A$, the fractions are ranked so that
\begin{equation}
 p_1(A;N)\geq p_2(A;N)\geq\cdots\geq p_{K_A}(A;N).
 \label{eq:ranked_basins}
\end{equation}
The coexistence of several attractors does not however imply that their basins are of comparable sizes. It turns out that for the shifted Alladi-Erd{\H o}s map, the phase space is distributed very unevenly among the existing attractors. To demonstrate this, we fix $N=10^6$, and then for a given shift $A$, we iterate the map starting from each initial condition $n \in \{2, 3, 4, \dots, N\}$ until we reach an attractor.
We find an interesting result that the two largest basins absorb nearly all initial conditions, so the quantity of interest is (we suppress the dependence on $N$):
\begin{equation}
 D_2(A)=p_1(A)+p_2(A),
 \label{eq:D2}
\end{equation}

\begin{figure}
    \includegraphics[width=0.9\columnwidth]{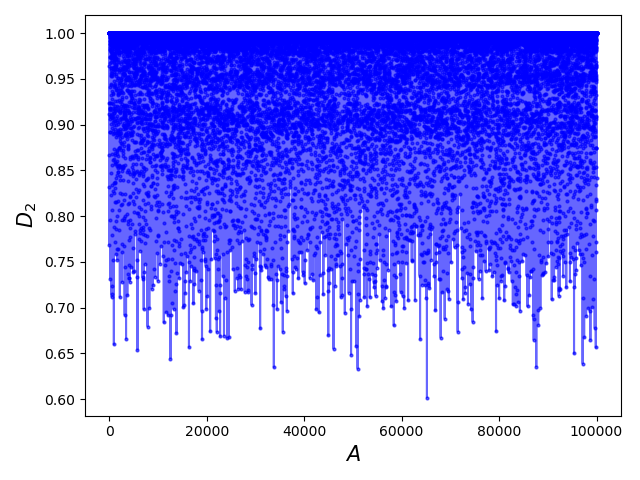}
    \caption{\label{fig:top_two_basin_sizes_vs_a_N1000000} Sum ($D_2=p_1+p_2$) of the two largest basin fractions as a function of the shift parameter $A \in [1, 10^5]$, where $p_1$ and $p_2$ denote the largest and second-largest basin fractions, respectively. Values of $D_2$ close to unity indicate that the two largest attractors together capture almost all initial conditions.}
\end{figure}

\begin{figure}
    \includegraphics[width=0.8\columnwidth]{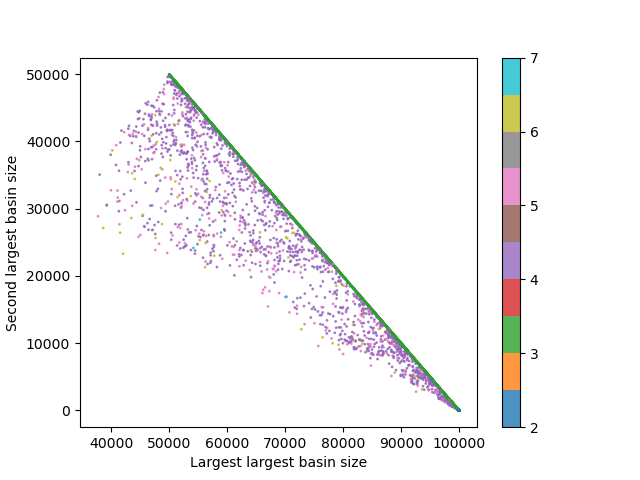}
    \caption{\label{fig:largest_second_largest_basin_size_scatter} Second-largest basin size plotted against the largest basin size for each $A \in [1, 10^5]$. The diagonal line represents $b_1+b_2=N-1$, where $N$ is the number of initial conditions used to estimate the basin sizes. Points close to this line correspond to systems in which the two largest basins together contain nearly all initial conditions. The color of each point indicates the number of coexisting attractors.}
\end{figure}

\begin{figure}
    \includegraphics[width=0.8\columnwidth]{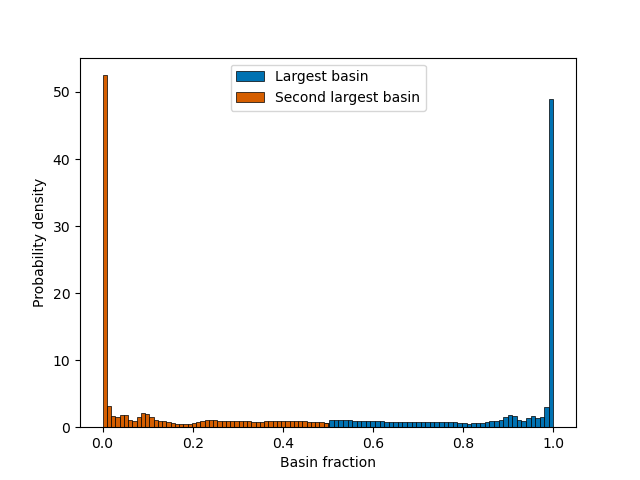}
    \caption{\label{fig:largest_second_largest_basin_size_hist} Probability-density histograms of the largest and second-largest basin fractions. The largest basin is typically close to the size of the full phase-space, whereas the second-largest basin usually has size zero since for many values of $A$ there is only one attractor as evident from Fig.~\ref{fig:num_cycles}(b). Bin edges were selected using the Freedman–Diaconis rule.}
\end{figure}

Fig.~\ref{fig:top_two_basin_sizes_vs_a_N1000000} shows the variation of $D_2$ as $A$ is varied from $1$ to $10^5$. Over this range, the lowest value of $D_2$ is around $0.6$, and its average value is $0.9168$ meaning that almost all initial conditions are captured by just two of the several attractors.
It is also interesting to see how the initial conditions are shared among the top two attractors. For this, we plot the scatterplot of their sizes shown in Fig.~\ref{fig:largest_second_largest_basin_size_scatter}. It is straightforward to see that since all basin sizes sum to $N-1$, the points must lie below the boundary $b_1+b_2=N-1$. The concentration of points near this line indicates that the first two basins almost exhaust the phase space for a large fraction of values of $A$. Finally, Fig.~\ref{fig:largest_second_largest_basin_size_hist} shows the histograms of the top two basin sizes obtained by varying $A$ over the range $[1, 10^5]$. 

\begin{figure}
    \includegraphics[width=0.9\columnwidth]{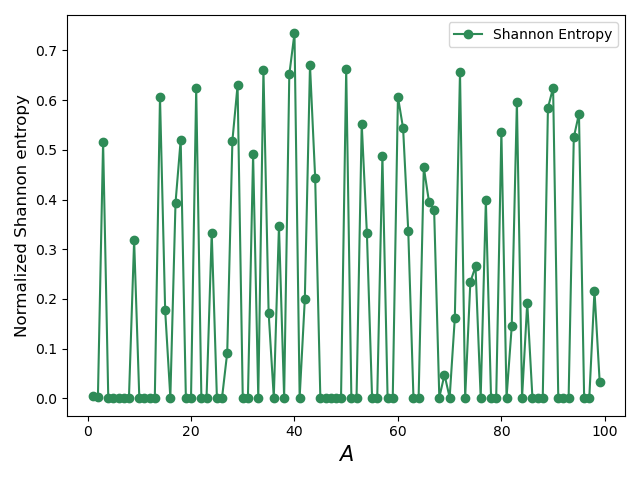}
    \caption{\label{fig:entropy_vs_a} Normalized basin entropy as a function of the shift parameter for $1\leq A\leq100$. A value close to zero indicates strong dominance by one or a few basins, whereas a value close to unity corresponds to basins of approximately equal size.}
\end{figure}

It is also possible to summarize the total number of attractors and sizes of their basins, by computing the Shannon entropy of the basins fractions as expressed by:
\begin{equation}
 H(A)=-\sum_{j=1}^{K_A}p_j(A)\log_2 p_j(A).
 \label{eq:entropy}
\end{equation}
The entropy measures how evenly the initial conditions are divided among the coexisting basins. It vanishes when a single basin occupies the complete window and increases as the basin fractions become more comparable. An even better quantity for this purpose is the normalized entropy obtained by dividing the Shannon entropy by the logarithm of the number of attractors:
\begin{equation}
 H_{\mathrm{norm}}(A)=\frac{H(A)}{\log_2 K_A},
 \label{eq:normalized_entropy}
\end{equation}
for $K_A>1$, and set $H_{\mathrm{norm}}=0$ when only one nontrivial basin is present (see Fig.~\ref{fig:entropy_vs_a}). 
We note that the use of entropy here is reminiscent of, but distinct from, the basin entropy introduced in \cite{daza2016basin} since there the phase space is partitioned into boxes and then local entropies of labels within boxes are averaged to get the basin entropy; here instead we are using Shannon entropy to summarize the number of attractors and sizes of their basins.

\section{Discussion and conclusions}
\label{sec:discussion}

We have revisited the shifted Alladi-Erd{\H o}s map to explore its rich structure for a significantly larger range of the shift parameter, and for the first time looked into its basin structure. In spite of its apparent simplicity, the map generates a nontrivial dynamics which provides a way to understand how multiplicative and additive structures in natural numbers interact with each other. In particular, the shift parameter $A$ probes these structures at different ``scales'': composite numbers descend through a fixed collection of prime trees, while the shift sends each prime root to another tree or to a composite number in the same tree. This decomposition is not just a useful computational trick. Rather, it is a the natural conceptual description of the system since varying $A$ only changes the connections among fixed trees rather than changing the trees themselves.

We showed that the smallest element of every nontrivial attractor of the map is a prime, and that its outgoing shifted step encodes $A$. Thus, a different set of attractor emerges for each value of $A$, and no nontrivial attractor is shared among different $A$ values. The situation differs from classical bifurcation theory, where the same invariant set can often be continued through a real parameter until it changes stability. Here individual cycles undergo complete turnover.

The compressed attractor diagram in Fig.~\ref{fig:compressed} nevertheless shows that for a given shift $A$, maps with nearby values of $A$ are not unrelated: all numbers in the cycles form line-like families as we have proved. 

The basin results reveal a second kind of large-scale regularity. Several attractors can coexist, but their dynamical importance is usually highly unequal. 
Basins of the largest and the second largest attractors fill more than $91\%$ phase space.
This leaves only a small residual fraction for all other attractors. The quantity $D_2=p_1+p_2$ expresses this hierarchy directly, while entropy quantifies the remaining diversity of the basin partition.

The work opens several questions. Can attractors corresponding to different values of $A$ be grouped into classes so that attractors in a given class are ``similar"? Is there a regularity in the sequence of primes and composites in these attractors? Is it possible to identify the complete basin of a given attractor analytically? For a fixed $A$, do basin sizes stabilize in the limit $N\to\infty$? Which arithmetic properties distinguish shifts with unusually large residual basin mass or high normalized entropy? What are the bounds on the slopes of the lines in the bifurcation diagram? The shifted Alladi-Erd{\H o}s map offers a simple setting in which such questions can be studied through a combination of number theory and discrete dynamics.

\appendix

\section{Numerical method and validation}
\label{app:numerics}

The computation exploits the separation between the fixed prime trees and the shift-dependent dynamics of their roots. To make sure that all the initial conditions up to a given $N$ are included, we constructed a smallest-prime-factor sieve up to $p_{max}+A_{max}$ where $p_{max}$ is the largest prime less than or equal to than $N$, and $A_{max}$ is the largest value of $A$ that is used for the exploration. This bound makes sure that every integer needed for a given range of shifts has been included in the sieve. Construction of the sieve permits rapid evaluation of the prime-factor sum $B(n)$ in Eq.~\eqref{eq:B} using recursion. 

For a given composite $n$, we also keep track of how many other composite numbers reach it through the repeated action of $B$. This ``weight" of $n$ can then be used to update the weight of the number $B(n)$. This speeds up basin size computation for attractors without full iteration for each number. 


For a fixed shift, the iteration is then carried out only on prime roots. From a prime $p$, if the successor $p+A$ is itself a prime, we keep repeating until we get a composite which can be mapped to the root of its prime tree. Once we land on the corresponding attractor cycle, we label all the numbers, starting from the initial one, using an index assigned to that attractor. The resulting functional graph is explored with memoization: a path is followed until it reaches a previously labeled root or repeats a root already on the current path (in which case we have a new attractor). The corresponding full cycle of $B_A$ is reconstructed by inserting the stored composite descent from each shifted value $p+A$ to the next prime root.

Once every relevant root has been assigned to an attractor, its precomputed tree weight is added to that attractor's basin count. Thus basin sizes are obtained without iterating every initial condition separately for every shift and without storing a complete basin-label vector for all values of $A$. For each shift, only summary data such as the canonical attractor cycles, cycle lengths, basin counts, and selected transient statistics are retained.

The main computations use $1\leq A\leq10^5$ and $2\leq n\leq10^6$. The sieve and root table must extend beyond $N$ whenever shifted values leave the initial-condition window as explained before.

Finite-window effects are assessed by repeating representative parts of the sweep for several values of $N$. Attractor identities and cycle lengths stabilize once all recurrent states have been included. Basin fractions converge more slowly because enlarging $N$ adds further levels to the prime trees. 

\section{Full attractor diagram}
\label{app:full_diagram}

\begin{figure}[ht]
    \centering
    \includegraphics[width=0.8\columnwidth]{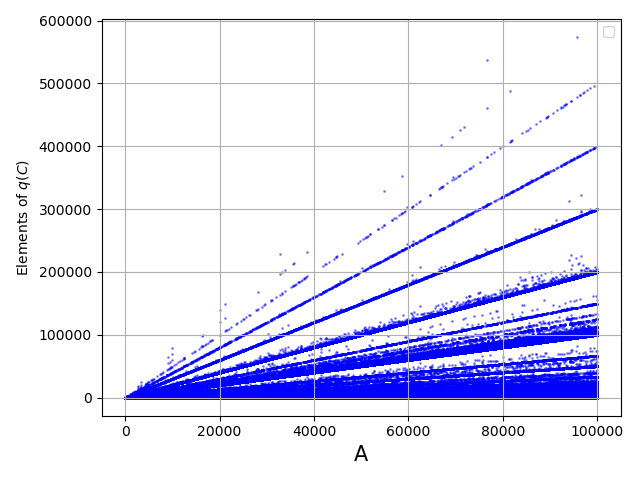}
    \caption{\label{fig:full_bifurcation}
    Conventional attractor diagram for shifts. Every number belonging to an attractor is plotted against $A$. Several branches are visible, but the lower cloud and substantial overplotting make the structure less legible than in the compressed representation of Fig.~\ref{fig:compressed}.}
\end{figure}
Figure~\ref{fig:full_bifurcation} shows the conventional diagram in which every element of every attractor is plotted. It records more information than Fig.~\ref{fig:compressed}, but the coexistence of many cycles and their broad numerical range produce substantial cluttering. The comparison motivates the use of the minimum-prime representation in the main text.

%

\end{document}